\documentclass[runningheads,a4paper]{llncs}

\usepackage{amssymb}
\providecommand{\myboldg}[1]    {\mbox{\boldmath $#1$}}         % lower greek
\providecommand{\mybolda}       {\myboldg}                      % lower alpha
\providecommand{\myboldA}       {\mybolda}                      % upper alpha
\providecommand{\myboldn}       {\mybolda}                      % numbers

\providecommand{\Bone}  {\myboldn{1}}

\providecommand{\bH}{\myboldA{H}}

\providecommand{\bM}{\myboldA{M}}

\providecommand{\bQ}{\myboldA{Q}}

\providecommand{\bS}{\myboldA{S}}

\providecommand{\bU}{\myboldA{U}}
\providecommand{\bV}{\myboldA{V}}
\providecommand{\bW}{\myboldA{W}}
\providecommand{\bX}{\myboldA{X}}
\providecommand{\bY}{\myboldA{Y}}
\providecommand{\bZ}{\myboldA{Z}}

\providecommand{\Balpha}        {\myboldg{\alpha}}

\usepackage{amsmath,graphicx}

\usepackage{multicol}
\usepackage[boxsize=2.3em]{ytableau}
\usepackage{sidecap}
\usepackage{color}
\usepackage{url}

\usepackage{url}
\urldef{\mailsa}\path|{grais, g.roma, andrew.simpson, m.plumbley}@surrey.ac.uk|    
\newcommand{\keywords}[1]{\par\addvspace\baselineskip
\noindent\keywordname\enspace\ignorespaces#1}

\usepackage{subcaption}
\begin{document}
\small

\mainmatter  % start of an individual contribution

\title{Discriminative Enhancement for Single Channel Audio Source Separation using Deep Neural Networks}

\titlerunning{Discriminative enhancement for SCSS using DNNs}

% Centre for Vision, Speech and Signal Processing, University of Surrey, Guildford, UK.
\author{Emad M. Grais \and Gerard Roma \and Andrew J.R. Simpson \and Mark D. Plumbley}
\authorrunning{Emad M. Grais et al.}

\institute{Centre for Vision, Speech and Signal Processing, University of Surrey, Guildford, UK.\\
\mailsa\\
%\mailsd\\
%\url{http://www.springer.com/lncs}
}


\toctitle{Lecture Notes in Computer Science}
\tocauthor{Authors' Instructions}
\maketitle

\begin{abstract}
The sources separated by most single channel audio source separation techniques are usually distorted and each separated source contains residual signals from the other sources. To tackle this problem, we propose to enhance the separated sources to decrease the distortion and interference between the separated sources using deep neural networks (DNNs). Two different DNNs are used in this work. The first DNN is used to separate the sources from the mixed signal. The second DNN is used to enhance the separated signals. To consider the interactions between the separated sources, we propose to use a single DNN to enhance all the separated sources together. To reduce the residual signals of one source from the other separated sources (interference), we train the DNN for enhancement discriminatively to maximize the dissimilarity between the predicted sources. The experimental results show that using discriminative enhancement decreases the distortion and interference between the separated sources.

\keywords{Single channel audio source separation, deep neural networks, audio enhancement, discriminative training.}
\end{abstract}

\section{Introduction}
%Single channel audio source separation (SCSS) aims to separate sources from their single mixture \cite{Virtanen:07:msssbnmfwtcasc,emad:12:hmmprnmfscss}. Various techniques based on nonnegative matrix factorization (NMF), Hidden Markov models, Gaussian mixture models, and different dictionary learning methods have been used to tackle this problem \cite{Radfar:06:anlmmseefma,Smaragdis:09:anafscsoks,grais:14:ssrnmfmmse}. 
Audio single channel source separation (SCSS) aims to separate sources from their single mixture \cite{emad:12:hmmprnmfscss,Virtanen:07:msssbnmfwtcasc}. Deep neural networks (DNNs) have recently been used to tackle the SCSS problem \cite{Erdogan:15:psrbssdrnn,Huang:14:svsmrdrnn,Felix:14:dtrnnscss,Williamson:14:tsaipqss}. DNNs have achieved better separation results than nonnegative matrix factorization which is considered as one of the most common approaches for the SCSS problem \cite{Erdogan:15:psrbssdrnn,Grais:14:dnnscss,Simpson:15:dkevmmcdnn,Felix:14:dtrnnscss}.  %\cite{Virtanen:07:msssbnmfwtcasc,emad:12:gmgprnmfscss,grais:14:ssrnmfmmse}. Training data for the sources are used to train the DNNs for separation.  
DNNs are used for SCSS to either predict the sources from the observed mixed signal \cite{Grais:14:dnnscss,Huang:14:svsmrdrnn}, or to predict time-frequency masks that are able to describe the contribution of each source in the mixed signal \cite{Erdogan:15:psrbssdrnn,Simpson:15:dkevmmcdnn,Felix:14:dtrnnscss}. The masks usually take bounded values between zero and one. It is normally preferred to train the DNNs to predict masks that take bounded values to avoid training them on the full dynamic ranges of the sources \cite{Erdogan:15:psrbssdrnn,Felix:14:dtrnnscss}. 

Most SCSS techniques produce separated sources accompanied by distortion and interference from other sources \cite{Erdogan:15:psrbssdrnn,emad:12:hmmprnmfscss,ozerov:09:fshmmfparass,Virtanen:07:msssbnmfwtcasc}. To improve the quality of the separated sources, Williamson et al.  \cite{Williamson:14:tsaipqss} proposed to enhance the separated sources using nonnegative matrix factorization (NMF). The training data for each source is modelled separately, and each separated source is enhanced individually by its own trained model. However, enhancing each separated source individually does not consider the interaction between the sources in the mixed signal \cite{emad:13:stpemenbscss,Williamson:14:tsaipqss}. Furthermore, the residuals of each source that appear in the other separated sources are not available to enhance their corresponding separated sources.%, which means, the remaining parts of each signal that appear in the other separated sources are lost.  

In this paper, to consider the interaction between the separated sources, we propose to enhance all the separated sources together using a single DNN. Using a single model to enhance all the separated sources together allows each separated source to be enhanced using its remaining parts that appear in the other separated sources. This means most of the available information of each source in the mixed signal can be used to enhance its corresponding separated source. DNNs have shown better performance than NMF in many audio signal enhancement applications \cite{Erdogan:15:psrbssdrnn,Felix:14:dtrnnscss}. Thus, in this work we use a DNN to enhance the separated sources rather than using NMF \cite{Williamson:14:tsaipqss}. We train the DNN for enhancement discriminatively to maximize the differences between the estimated sources \cite{Huang:14:svsmrdrnn,Huang:15:jomdrnnmss}. A new cost function to discriminatively train the DNN for enhancement is introduced in this work. Discriminative training for the DNN aims to decrease the interference of each source in the other estimated sources and has also been found to decrease distortions \cite{Huang:14:svsmrdrnn}. Unlike other enhancement approaches such as NMF \cite{Williamson:14:tsaipqss} and denoising deep autoencoders \cite{Vincent:10:sdalurdnldc,Xie:12:ididnn} that aim to only enhance the quality of an individual signal, our new discriminative enhancement approach in this work aims to both enhance the quality and achieve good separation for the estimated sources. %In this work, we use a DNN for separating the mixed signal. The DNN for separation in this work is used to predict a mask that separates the sources from the mixed signal. After separating the sources, another DNN is used to enhance all the separated sources together. 

The main contributions of this paper are: (1) the use of a single DNN to enhance all the separated signals together; (2) discriminative training of a DNN for enhancing the separated sources to maximize the dissimilarity of the predicted sources; (3) a new cost function for discriminatively training the DNN.

This paper is organized as follows. In Section \ref{pf} a mathematical formulation of the SCSS problem is given. Section \ref{sec:DNNS} presents our proposed approach for using DNNs for source separation and enhancement. The experimental results and the conclusion of this paper are presented in Sections \ref{sec:exp} and \ref{sec:conc}. 

\section{Problem formulation of audio SCSS}
\label{pf}
Given a mixture of $I$ sources as $y(t) = \sum_{i=1}^I s_i(t)$, the aim of audio SCSS is to estimate the sources $s_i(t), \ \forall{i}$, from the mixed signal $y(t)$. %This can be formulated in the short time Fourier transform (STFT) domain as $Y(n,f)=\sum_{i=1}^I S_i(n,f)$, where $S_i(n,f)$ is the unknown STFT of source $s_i(t)$ and $Y(n,f)$ in the STFT of the observed mixed signal $y(t)$. 
The estimate $\hat{S}_i(n,f)$ for source $i$ in the short time Fourier transform (STFT) domain can be found by predicting a time-frequency mask $M_i(n,f)$ that scales the mixed signal according to the contribution of source $i$ in the mixed signal as follows \cite{Erdogan:15:psrbssdrnn,Simpson:15:dkevmmcdnn,Felix:14:dtrnnscss}: 
\begin{equation}
\label{est_trgt1_pf}
%\footnotesize
%{
\hat {S}_i(n,f)=M_i(n,f) \times Y(n,f)
%}
\end{equation}
where $Y(n,f)$ is the STFT of the observed mixed signal $y(t)$, while $n$ and $f$ are the time and frequency indices respectively. The mask $M_i(n,f)$ takes real values between zero and one. The main goal here is to predict masks $M_i(n,f), \ \forall{i}$, that separate the sources from the mixed signal. In this framework, the magnitude spectrogram of the mixed signal is approximated as a sum of the magnitude spectra of the estimated sources \cite{ozerov:09:fshmmfparass,Virtanen:07:msssbnmfwtcasc} as follows:
\begin{equation}
\label{stftangl_2}
\left|Y(n,f)\right|\approx \sum_{i=1}^I\left| \hat{S}_i(n,f)\right|.
\end{equation}
For the rest of this paper, we denote the magnitude spectrograms and the masks in a matrix form as $\bY$, $\hat \bS_i$, and $\bM_i$.% after dropping the indices $n$ and $f$. 
%and 
%\begin{equation}
%\label{est_trgt2}
%\hat {\bS}_2 = \left(\Bone-\bM\right) \odot \bY,
%\end{equation}
%where $\odot$ represents an elementwise multiplication and $\Bone$ is a matrix of ones.

\section{DNN\small{s} for source separation and enhancement}
\label{sec:DNNS}
\begin{SCfigure}
\centering
\includegraphics[width=0.50\linewidth]{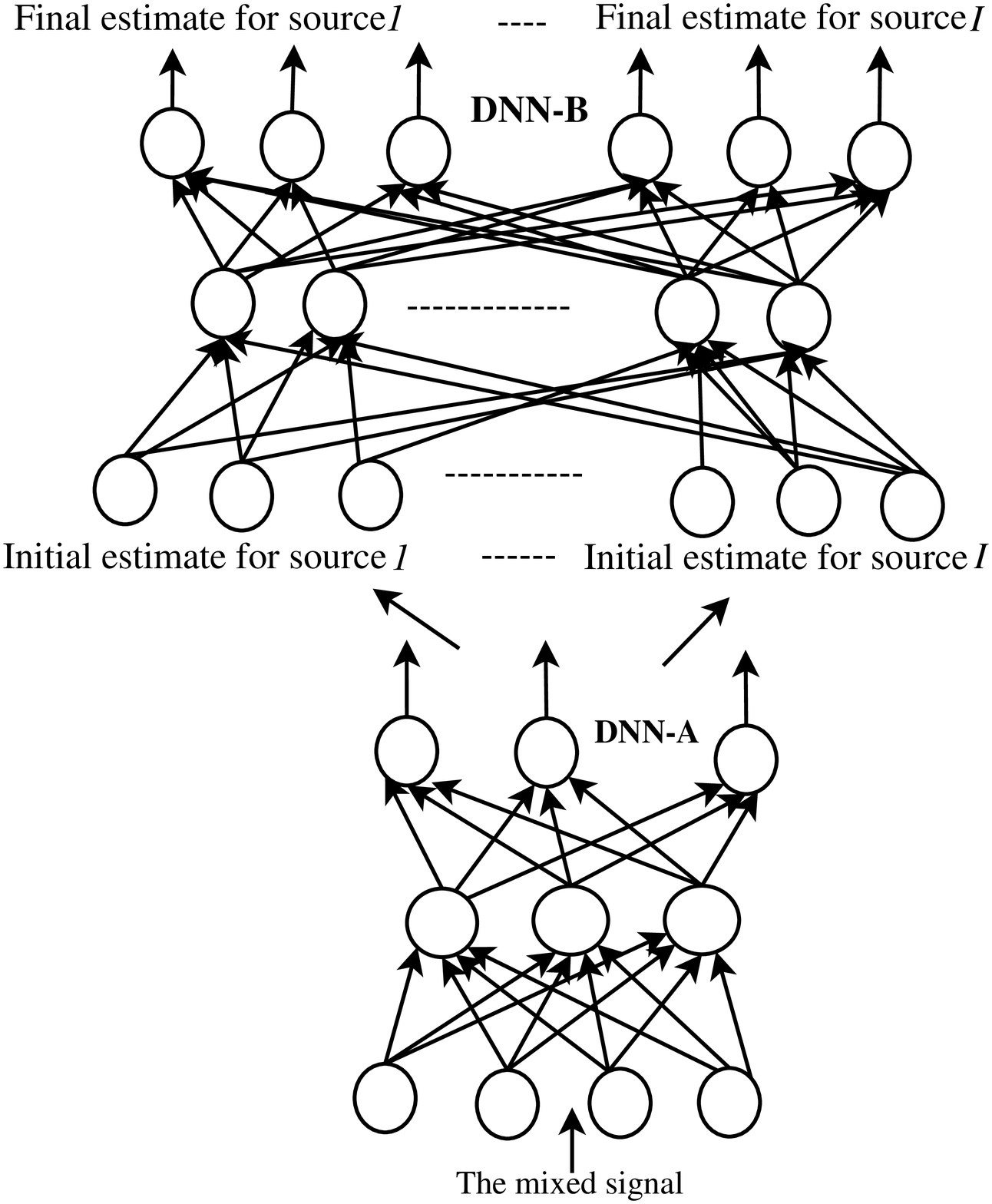}%chart1_email
\caption{\label{fig:dnn1}The overview of the proposed approach of using DNNs for source separation and enhancement. DNN-A is used for separation. DNN-B is used for enhancement.}
%\end{center}
\end{SCfigure}

In this paper, we use two deep neural networks (DNNs) to perform source separation and enhancement. The first DNN (DNN-A) is used to separate the sources from the mixed signal. The separated sources are then enhanced by the second DNN (DNN-B) as shown in Figure \ref{fig:dnn1}. DNN-A is trained to map the mixed signal in its input into reference masks in its output.  
%The reference masks are computed based on the ratio between the sources in each time-frequency bin of the magnitude spectrogram of the mixed signal. 
DNN-B is trained to map the separated sources (distorted signals) from DNN-A into their reference/clean signals. As in many machine learning tasks, the data used to train the DNNs is usually different than the data used for testing \cite{Erdogan:15:psrbssdrnn,Huang:14:svsmrdrnn,Simpson:15:dkevmmcdnn,Felix:14:dtrnnscss}. The performance of the trained DNN on the test data is often worse than the performance on the training set. The trained DNN-A is used to separate data that is different than the training data, and since the main goal of using DNN-B is to enhance the separated signals by DNN-A, then DNN-B should be trained on a different set of data than the set of data that was used to train DNN-A. Thus, in this work we divide the available training data into two sets. The first set of the training data is used to train DNN-A for separation and the second set is for training DNN-B for enhancement. %The proposed procedures for training and testing the DNNs for separation and enhancement are shown in the following subsections.

\subsection{Training DNN-A for source separation}
\label{subsec:trn_dnn_scss}
Given the magnitude spectrograms of the sources in the first set of the training data $\bS_{\scriptsize{\mbox{tr}}i}^{(1)}, \ \forall i$, DNN-A is trained to predict a reference mask $\bM_{\scriptsize{\mbox{tr}}i}^{(1)}$. The subscript $\scriptsize{\mbox{tr}i}$ indicates the training data for source $i$, and the superscript ``(1)'' indicates the first set of the training data is used for training. Different types of masks have been proposed in \cite{Narayanan:13:irmednnrsr,Felix:14:dtrnnscss}. We chose to use the ratio mask from \cite{Felix:14:dtrnnscss}, which  gives separated sources with reasonable distortion and interference. The reference ratio mask in \cite{Felix:14:dtrnnscss} is defined as follows:
\begin{equation}
\label{soft_mask}
\bM_{\scriptsize{\mbox{tr}}i}^{(1)} = \frac{\bS_{\scriptsize{\mbox{tr}}i}^{(1)}}{\sum_{i=1}^I\bS_{\scriptsize{\mbox{tr}}i}^{(1)}}
\end{equation}
where the division is done element-wise, $\bS_{\scriptsize{\mbox{tr}}i}^{(1)}$ is the magnitude spectrogram of reference source $i$, and $\bM_{\scriptsize{\mbox{tr}}i}^{(1)}$ is the mask which defines the contribution of source $i$ in every time-frequency bin ($n,f$) in the mixed signal. The input of DNN-A is the magnitude spectrogram $\bX^{(1)}_{\scriptsize{\mbox{tr}}}$ of the mixed signal of the first set of the training data which is formulated as $\bX^{(1)}_{\scriptsize{\mbox{tr}}}=\sum_{i=1}^I\bS_{\scriptsize{\mbox{tr}}i}^{(1)}$. The reference/target output of DNN-A for all sources is formed by concatenating the reference masks for all sources as 
\begin{equation}
\label{total_mask}
\bM_{\scriptsize{\mbox{tr}}}^{(1)} = \left[\bM_{\scriptsize{\mbox{tr}}1}^{(1)},\ldots,\bM_{\scriptsize{\mbox{tr}}i}^{(1)}, \ldots,\bM_{\scriptsize{\mbox{tr}}{I}}^{(1)}\right]. 
\end{equation}
%Since the last mask $\bM_{\scriptsize{\mbox{tr}}I}^{(1)}$ can be computed as $\bM_{\scriptsize{\mbox{tr}}I}^{(1)}=\Bone - \sum_{i=1}^{I-1} \bM_{\scriptsize{\mbox{tr}}i}^{(1)}$ where $\Bone$ is a matrix of ones, then there is no need to train DNN-A to predict $\bM_{\scriptsize{\mbox{tr}}I}^{(1)}$. 
DNN-A is trained to minimize the following cost function as in \cite{arie:16:masswdnn,Felix:14:dtrnnscss}:
\begin{equation}
\label{cost_mask}
C_1 =\sum_{n,f}\left( \bZ_{\scriptsize{\mbox{tr}}}^{(1)}\left(n,f\right) - \bM_{\scriptsize{\mbox{tr}}}^{(1)}\left(n,f\right) \right)^2
\end{equation}
where $\bZ_{\scriptsize{\mbox{tr}}}^{(1)}$ is the actual output of the final layer of DNN-A and $\bM_{\scriptsize{\mbox{tr}}}^{(1)}\in \left[0,1\right]$ is computed from Eqs. (\ref{soft_mask}) and (\ref{total_mask}). The activation functions of the output layer for DNN-A are sigmoid functions, thus $\bZ_{\scriptsize{\mbox{tr}}}^{(1)}\in \left[0,1\right]$.  

\subsection{Training DNN-B for discriminative enhancement}
\label{subsec:trn_dnn_enhs}
%The main idea of using the second DNN2 is to enhance the distorted separated signals that are separated by DNN1. DNN2 should be trained using data that is separated by DNN1. The data that DNN1 separates in this stage should not be the same data that was used for training DNN1, which is a practical scenario for using DNNs for SCSS \cite{Erdogan:15:psrbssdrnn,Williamson:14:tsaipqss}. In practical situations, DNNs are usually used to separate data that is different than the set of data that was used for training \cite{Huang:14:svsmrdrnn,Felix:14:dtrnnscss}. 
To generate the training data to train DNN-B, the trained DNN-A is used to separate mixed signals from the second set of the training data. The mixed signal of this set of training data is formulated as $\bX_{\scriptsize{\mbox{tr}}}^{(2)}=\sum_{i=1}^I\bS_{\scriptsize{\mbox{tr}}i}^{(2)}$, where $\bX_{\scriptsize{\mbox{tr}}}^{(2)}$ is the magnitude spectrogram of the mixed signal in the second set of the training data, the superscript ``(2)'' indicates that the second set of the training data is used in this stage. The frames of $\bX_{\scriptsize{\mbox{tr}}}^{(2)}$ are fed as inputs to DNN-A, which then produces mask $\bZ_{\scriptsize{\mbox{tr}}}^{(2)}$ which is a concatenation of masks for many sources as $\bZ_{\scriptsize{\mbox{tr}}}^{(2)} = \left[\bZ_{\scriptsize{\mbox{tr}}1}^{(2)},\ldots,\bZ_{\scriptsize{\mbox{tr}}i}^{(2)}, \ldots,\bZ_{\scriptsize{\mbox{tr}}{I}}^{(2)}\right]$. %The mask for the last source is estimated as $\bZ_{\scriptsize{\mbox{tr}}I}^{(2)}=\Bone - \sum_{i=1}^{I-1} \bZ_{\scriptsize{\mbox{tr}}i}^{(2)}$. 
The estimated masks are used to estimate the sources as follows:
\begin{equation}
\label{est_trgt1_tr}
\tilde {\bS}_{\scriptsize{\mbox{tr}}i}^{(2)} = \bZ_{\scriptsize{\mbox{tr}}i}^{(2)} \odot \bX_{\scriptsize{\mbox{tr}}}^{(2)}, \forall i 
\end{equation}
%and the estimate for the second source is
%\begin{equation}
%\label{est_trgt2}
%\hat {\bS}_{tr2}^{(2)}= \left(\Bone - \bZ_2 \right) \odot \bX^{(2)}.
%\end{equation}
where $\odot$ denotes an element-wise multiplication. Each separated source $\tilde {\bS}_{\scriptsize{\mbox{tr}}i}^{(2)}$ often contains remaining signals from the other sources.  
%Since $\bX_{\scriptsize{\mbox{tr}}}^{(2)} =  \sum_{i=1}^I \tilde{\bS}_{\scriptsize{\mbox{tr}}i}^{(2)}$, the data in $\bX_{\scriptsize{\mbox{tr}}}^{(2)}$ is still preserved in the estimated sources $\tilde {\bS}_{\scriptsize{\mbox{tr}}i}^{(2)}, \forall i$. If some data are missing in the estimate $\tilde {\bS}_{\scriptsize{\mbox{tr}}i}^{(2)}$ comparing with its reference signal $\bS_{\scriptsize{\mbox{tr}}i}^{(2)}$, these missing data will appear in the estimates for the other sources $\tilde {\bS}_{\scriptsize{\mbox{tr}}{j}}^{(2)}, \ \forall j \neq i$. 
In this work, to consider the available information of each source that appears in the other separated sources, we propose to train DNN-B to enhance all the separated sources $\tilde {\bS}_{\scriptsize{\mbox{tr}}i}^{(2)}, \forall i$ together. %This means, every separated source may have parts in the other separated sources and to enhance each source, 

DNN-B is trained using the separated signals $\tilde {\bS}_{\scriptsize{\mbox{tr}}i}^{(2)}, \forall i$ and their corresponding reference/clean signals $\bS_{\scriptsize{\mbox{tr}}i}^{(2)}, \forall i$. The input for DNN-B is the concatenation of the separated signals $\bU_{\scriptsize{\mbox{tr}}}^{(2)} =\left[\tilde {\bS}_{\scriptsize{\mbox{tr}}1}^{(2)}, \ldots, \tilde {\bS}_{\scriptsize{\mbox{tr}}i}^{(2)}, \ldots, \tilde {\bS}_{\scriptsize{\mbox{tr}}I}^{(2)}\right]$. DNN-B is trained to produce in its output layer the concatenation of the reference signals as $\bV_{\scriptsize{\mbox{tr}}}^{(2)}=\left[\bS_{\scriptsize{\mbox{tr}}1}^{(2)}, \ldots,  \bS_{\scriptsize{\mbox{tr}}i}^{(2)}, \ldots, \bS_{\scriptsize{\mbox{tr}}I}^{(2)}\right]$. Each frame in $\bS_{\scriptsize{\mbox{tr}}i}^{(2)}, \forall i$ is normalized to have a unit Euclidean norm. This normalization allows us to train DNN-B to produce bounded values in its output layer without any need to train DNN-B over a wide range of values that the sources can have. Since the reference normalized signals have values between zero and one, we choose the activation functions of the output layer of DNN-B to be sigmoid functions. %Since every input and output frames of DNN-B is a concatenation of all sources, the dimension of the input and output vectors for the DNN-B is twice the dimension of the input of DNN1 as shown in Fig. \ref{fig:dnn1}.
% DNN2 is wider than DNN1.
%
%\begin{figure}[h]
%\begin{center}
%\includegraphics[width=1\linewidth]{Enhance_2-eps-converted-to}%chart1_email
%\caption{\label{fig:dnn2}Illustration of DNN2 architecture.}
%\end{center}
%\end{figure}

DNN-B is trained to minimize the following proposed cost function:
\begin{equation}
%\begin{multline}
%\begin{split}
\label{cost_sgn_discr}
C_2  =\sum_{n,f}\left(\bQ_{\scriptsize{\mbox{tr}}}^{(2)}\left(n,f\right) - \bV_{\scriptsize{\mbox{tr}}}^{(2)}\left(n,f\right)\right)^2 
        - \lambda \sum_{j \neq i}^I \sum_{n,f}\left(\bQ_{\scriptsize{\mbox{tr}}i}^{(2)}\left(n,f\right) - \bS_{\scriptsize{\mbox{tr}}{j}}^{(2)}\left(n,f\right)\right)^2
       %\end{multline}
%\end{split}
\end{equation}
where $\lambda$ is a regularization parameter, $\bQ_{\scriptsize{\mbox{tr}}}^{(2)}$ is the actual output of DNN-B which is a concatenation of estimates for all sources as $\bQ_{\scriptsize{\mbox{tr}}}^{(2)}=\left[\bQ_{\scriptsize{\mbox{tr}}1}^{(2)}, \ldots, \bQ_{\scriptsize{\mbox{tr}}i}^{(2)}, \ldots, \bQ_{\scriptsize{\mbox{tr}}I}^{(2)}\right]$. The output $\bQ_{\scriptsize{\mbox{tr}}i}^{(2)}$ is the set of DNN-B output nodes that correspond to the normalized reference output $\bS_{\scriptsize{\mbox{tr}}i}^{(2)}$. The first term in the cost function in Eq. (\ref{cost_sgn_discr}) minimizes the difference between the outputs of DNN-B and their corresponding reference signals. The second term of the cost function maximizes the dissimilarity/differences between DNN-B outputs of different sources, which is considered as ``discriminative learning" \cite{Huang:14:svsmrdrnn,Huang:15:jomdrnnmss}. The cost function in Eq. (\ref{cost_sgn_discr}) aims to decrease the possibility of each set of the outputs of DNN-B from representing the other set, which helps in achieving better separation for the estimated sources. Note that, DNN-A is trained to predict masks in its output layer, while DNN-B is trained to predict normalized magnitude spectrograms for the sources. Both DNNs are trained to produce bounded values between zero and one.     

\subsection{Testing DNN-A and DNN-B}
\label{subsec:test_dnn_enhs}
In the separation stage, we aim to use the trained DNNs (DNN-A and DNN-B) to separate the sources from the mixed signal. Given the magnitude spectrogram $\bY$ of the mixed signal $y(t)$. The frames of $\bY$ are fed to DNN-A to predict concatenated masks in its output layer as $\tilde {\bZ}_{\scriptsize{\mbox{ts}}} = \left[\tilde {\bZ}_{\scriptsize{\mbox{ts}}1},\ldots,\tilde {\bZ}_{\scriptsize{\mbox{ts}}i}, \ldots,\tilde {\bZ}_{\scriptsize{\mbox{ts}}{I}}\right]$. 
%The mask for the last source is then computed as $\tilde {\bZ}_{\scriptsize{\mbox{ts}}I}=\Bone - \sum_{i=1}^{I-1} \tilde {\bZ}_{\scriptsize{\mbox{ts}}i}$. 
The output masks are then used to compute initial estimates for the magnitude spectra of the sources as follows:
%($\bZ_{\scriptsize{\mbox{ts}}}$). The masks $\bZ_{\scriptsize{\mbox{ts}}}$ is used to find initial estimates for the source signals $\tilde {\bS}_{\scriptsize{\mbox{ts1}}}$ and $\tilde {\bS}_{\scriptsize{\mbox{ts2}}}$ as follows:
\begin{equation}
\label{est_trgt1}
\tilde {\bS}_{\scriptsize{\mbox{ts}}i} = \bZ_{\scriptsize{\mbox{ts}}i} \odot \bY, \forall i. 
\end{equation}
The initial estimates for the sources $\tilde {\bS}_{\scriptsize{\mbox{ts}}i}$ are usually distorted \cite{emad:13:stpemenbscss,Williamson:14:tsaipqss}, and need to be enhanced by DNN-B. The sources can have any values but the output nodes of DNN-B are composed of sigmoid activation functions that take values between zero and one. To retain the scale information between the sources, the Euclidean norm (gain) of each frame in the spectrograms of the estimated source signals $\tilde {\bS}_{\scriptsize{\mbox{ts}}i}, \forall i$ are computed as $\Balpha_{\scriptsize{\mbox{ts}}i}=\left[ \alpha_{1,i},..,\alpha_{n,i},..,\alpha_{N,i} \right]$ and saved to be used later, where $N$ is the number of frames in each source. The estimated sources are concatenated as $\tilde {\bS}_{\scriptsize{\mbox{ts}}} = \left[\tilde {\bS}_{\scriptsize{\mbox{ts}}1}, \ldots, \tilde {\bS}_{\scriptsize{\mbox{ts}}i}, \ldots, \tilde {\bS}_{\scriptsize{\mbox{ts}}I}\right]$, and then fed to DNN-B to produce a concatenation of estimates for all sources $\hat {\bS}_{\scriptsize{\mbox{ts}}}=\left[\hat {\bS}_{\scriptsize{\mbox{ts}}1}, \ldots, \hat {\bS}_{\scriptsize{\mbox{ts}}i}, \ldots, \hat {\bS}_{\scriptsize{\mbox{ts}}I}\right]$. The values of the outputs of DNN-B are between zero and one. The output of DNN-B is then used with the gains in $\Balpha_{\scriptsize{\mbox{ts}}i}, \forall i$ to build a final mask as follows:

\begin{equation}
\label{est_trgt_final}
\bM_{\scriptsize{\mbox{ts}}i} = \frac{\Balpha_{\scriptsize{\mbox{ts}}i} \otimes \hat {\bS}_{\scriptsize{\mbox{ts}}i}}{\sum_{i=1}^I \Balpha_{\scriptsize{\mbox{ts}}i} \otimes \hat {\bS}_{\scriptsize{\mbox{ts}}i}}
\end{equation}
where the division here is also element-wise and the multiplication $\Balpha_{\scriptsize{\mbox{ts}}i} \otimes \hat {\bS}_{\scriptsize{\mbox{ts}}i}$ means that each frame $n$ in $\hat {\bS}_{\scriptsize{\mbox{ts}}i}$ is multiplied (scaled) with its corresponding gain entry $\alpha_{n,i}$ in $\Balpha_{\scriptsize{\mbox{ts}}i}$. The scaling using $\Balpha_{\scriptsize{\mbox{ts}}i}$ here helps in using DNN-B with bounded outputs between zero and one without the need to train DNN-B over all possible values of the source signals. Each $\alpha_{n,i}$ here is considered as an estimate for the scale of its corresponding frame $n$ in source $i$.        
The final enhanced estimate for the magnitude spectrogram of each source $i$ is computed as
\begin{equation}
\label{est_trgt1_fnl}
\hat {\bS}_i=\bM_{\scriptsize{\mbox{ts}}i} \odot \bY.
\end{equation} 
%Using the mask $\bM$ guarantees that the magnitude spectrograms of the enhanced sources sum up to the magnitude spectrogram of the mixed signal.   
The time domain estimate for source $\hat s_i(t)$ is computed using the inverse STFT of $\hat {\bS}_i$ with the phase angle of the STFT of the mixed signal. 

\section{Experiments and Discussion}
\label{sec:exp}
We applied the proposed separation and enhancement approaches to separate vocal and music signals from various songs in the dataset of SiSEC-2015-MUS-task \cite{ono:15:tsisec}. The dataset has 100 stereo songs with different genres and instrumentations. To use the data for the proposed SCSS approach, we converted the stereo songs into mono by computing the average of the two channels for all songs and sources in the data set. We consider to separate each song into vocal signals and accompaniment signals. The accompaniment signals tend to have higher energy than the vocal signals in most of the songs in this dataset \cite{ono:15:tsisec}.
The first 35 songs were used to train DNN-A for separation as shown in Section \ref{subsec:trn_dnn_scss}. The next 35 songs were used to train DNN-B for enhancement as shown in Section \ref{subsec:trn_dnn_enhs}. The remaining 30 songs were used for testing. The data was sampled at 44.1kHz. The magnitude spectrograms for the data were calculated using the STFT: a Hanning window with 2048 points length and overlap interval of 512 was used and the FFT was taken at 2048 points, the first 1025 FFT points only were used as features for the data.

%we have experimented with different values and we achieve reasonable results with the following parameters for the DNNs
For the parameters of the DNNs: For DNN-A, the number of nodes in each hidden layer was 1025 with three hidden layers. Since we separate two sources, DNN-A is trained to produce a single mask for the vocal signals $\bM_{\scriptsize{\mbox{voc}}}^{(1)}$ in its output layer and the second mask that separates the accompaniment source is computed as $\bM_{\scriptsize{\mbox{acc}}}^{(1)} = \Bone - \bM_{\scriptsize{\mbox{voc}}}^{(1)}$, where $\Bone$ is a matrix of ones. Thus, the dimension of the output layer of DNN-A is 1025.
For DNN-B, the number of nodes in the input and output layers is 2050 which is the length of the concatenation of the two sources $2 \times 1025$. For DNN-B, we used three hidden layers with 4100 nodes in each hidden layer. Sigmoid nonlinearity was used at each node including the output nodes for both DNNs. The parameters for the DNNs were initialized randomly. We used 200 epochs for backpropagation training for each DNN. Stochastic gradient descent was used with batch size 100 frames and learning rate $0.1$. We implemented our proposed algorithms using Theano \cite{Bergstra:10:theano2}. For the regularization parameter $\lambda$ in Eq. (\ref{cost_sgn_discr}), we tested with different values as shown in Fig. \ref{fig:sdrsir} below. We also show the results of using enhancement without discriminative learning where $\lambda=0$.

We compared our proposed discriminative enhancement approach using DNN with using NMF to enhance the separated signals similar to \cite{Williamson:14:tsaipqss}. In \cite{Williamson:14:tsaipqss}, a DNN was used to separate speech signals from different background noise signals and then NMF was used to improve the quality of the separated speech signals only. Here we modified the method in \cite{Williamson:14:tsaipqss} to suit the application of enhancing all the separated sources. 
%We compared our proposed DNN for enhancement approach in this work with using NMF for post-enhancement similar to \cite{Williamson:14:tsaipqss} with slight modifications to get better results than \cite{Williamson:14:tsaipqss}. 
NMF uses the magnitude spectrograms of the training data in Section \ref{subsec:trn_dnn_enhs} to train basis matrices $\bW_{\scriptsize{\mbox{tr1}}}$ and $\bW_{\scriptsize{\mbox{tr2}}}$ for both sources as follows:  
\begin{equation}
\label{NMF_train}
\bS_{\scriptsize{\mbox{tr1}}}^{(2)} \approx \bW_{\scriptsize{\mbox{tr1}}}\bH_{\scriptsize{\mbox{tr1}}} \ \ \mbox{and} \ \ \bS_{\scriptsize{\mbox{tr2}}}^{(2)} \approx \bW_{\scriptsize{\mbox{tr2}}}\bH_{\scriptsize{\mbox{tr2}}}
\end{equation} 
where $\bH_{\scriptsize{\mbox{tr1}}}$ and $\bH_{\scriptsize{\mbox{tr2}}}$ contain the gains of the basis vectors in $\bW_{\scriptsize{\mbox{tr1}}}$ and $\bW_{\scriptsize{\mbox{tr2}}}$ respectively. As in \cite{Williamson:14:tsaipqss}, we trained 80 basis vectors for each source and the generalized Kullback-Leibler divergence \cite{Lee:01:afnmf} was used as a cost function for NMF. NMF was then used to decompose the separated spectrograms $\tilde {\bS}_{\scriptsize{\mbox{ts}}i}, \forall i=1,2$ in Eq. (\ref{est_trgt1}) with the trained basis matrices $\bW_{\scriptsize{\mbox{tr1}}}$ and $\bW_{\scriptsize{\mbox{tr2}}}$ as follows: 
\begin{equation}
\label{NMF_test}
\tilde {\bS}_{\scriptsize{\mbox{ts1}}} \approx \bW_{\scriptsize{\mbox{tr1}}}\bH_{\scriptsize{\mbox{tst1}}} \ \  \mbox{and} \ \ \tilde {\bS}_{\scriptsize{\mbox{ts2}}} \approx \bW_{\scriptsize{\mbox{tr2}}}\bH_{\scriptsize{\mbox{tst2}}}
\end{equation}
where the gain matrices $\bH_{\scriptsize{\mbox{tst1}}}$ and $\bH_{\scriptsize{\mbox{tst2}}}$ contain the contribution of each trained basis vector of $\bW_{\scriptsize{\mbox{tr1}}}$ and $\bW_{\scriptsize{\mbox{tr2}}}$ in the mixed signal. In \cite{Williamson:14:tsaipqss}, the product $\bW_{\scriptsize{\mbox{tr1}}}\bH_{\scriptsize{\mbox{tst1}}}$ was used directly as an enhanced-separated speech signal. Here we used the product $\bW_{\scriptsize{\mbox{tr1}}}\bH_{\scriptsize{\mbox{tst1}}}$ and $\bW_{\scriptsize{\mbox{tr2}}}\bH_{\scriptsize{\mbox{tst2}}}$ to build a mask equivalent to Eq. (\ref{est_trgt_final}) as follows:
\begin{equation}
\label{est_trgt_final_nmf}
\bM_{1_{\mbox{nmf}}} = \frac{\bW_{\scriptsize{\mbox{tr1}}}\bH_{\scriptsize{\mbox{tst1}}}}{\bW_{\scriptsize{\mbox{tr1}}}\bH_{\scriptsize{\mbox{tst1}}} + \bW_{\scriptsize{\mbox{tr2}}}\bH_{\scriptsize{\mbox{tst2}}}}, \ \ \mbox{and} \ \ \bM_{2_{\mbox{nmf}}}= \Bone-\bM_{1_{\mbox{nmf}}}.
\end{equation}
These masks are then used to find the final estimates for the source signals as in Eq. (\ref{est_trgt1_fnl}). 

Performance of the separation and enhancement algorithms was measured using the signal to distortion ratio (SDR), signal to interference ratio (SIR), and signal to artefact ratio (SAR) \cite{vincent:06:pmi}. SIR indicates how well the sources are separated based on the remaining interference between the sources after separation. SAR indicates the artefacts caused by the separation algorithm to the estimated separated sources. SDR measures how distorted the separated sources are. The SDR values are usually considered as the overall performance evaluation for any source separation approach \cite{vincent:06:pmi}. Achieving high SDR, SIR, and SAR indicates good separation performance. 
\begin{figure}[h]
\begin{center}
\includegraphics[width=1\linewidth,height=6.5cm]{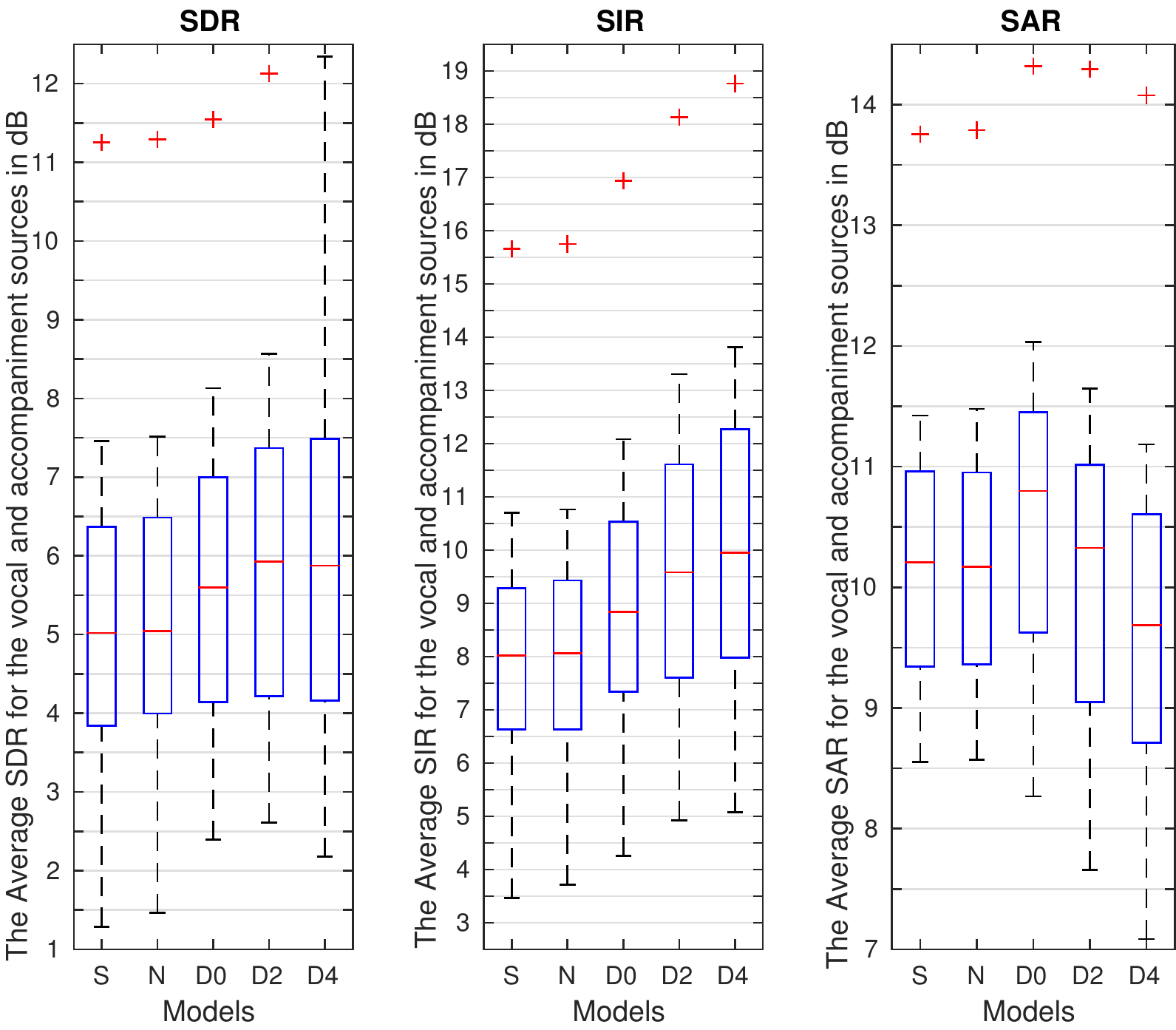}%chart1_email,keepaspectratio
\caption{\label{fig:sdrsir}The box-plot of the average SDR, SIR, and SAR of the vocal and accompaniment signals for the test set. Model ``S'' is for using DNN-A for source separation without enhancement. Model ``N'' is for using DNN-A for separation and NMF for enhancement. Models D0, D2, and D4 are for using DNN-A for separation followed by using DNN-B for enhancement with regularization parameter $\lambda=0.0,0.2$, and $0.4$ receptively.}
\end{center}
\end{figure}
\begin{table}[h]
\caption{The significant differences between each pair of models in Fig. \ref{fig:sdrsir}. the signs + and - in each cell at a certain row and column mean that the model in this row is significantly better or worse respectively than the model in this column, the sign ``0'' means no evidence for significant differences between the models. Model S is for separation only using DNN-A without enhancement. Models N, D0, D2, and D4 are for enhancing the separated sources using NMF, DNN-B with $\lambda=0$, DNN-B with $\lambda=0.2$, and DNN-B with $\lambda=0.4$ respectively.}
\begin{multicols}{3}
\scalebox{0.7}
{
\begin{ytableau}
\mbox{D4}&*(green!20)\mbox{+}&*(green!20)\mbox{+}&*(yellow!20)\mbox{0}&*(yellow!20)\mbox{0}& \\
\mbox{D2}&*(green!20)\mbox{+}&*(green!20)\mbox{+}&*(green!20)\mbox{+}&  &*(yellow!20)\mbox{0}\\
\mbox{D0}&*(green!20)\mbox{+}&*(green!20)\mbox{+}&  &*(red!20)\mbox{-}&*(yellow!20)\mbox{0}\\
\mbox{N}&*(green!20)\mbox{+}&  &*(red!20)\mbox{-}&*(red!20)\mbox{-}&*(red!20)\mbox{-}\\
\mbox{S}&  &*(red!20)\mbox{-}&*(red!20)\mbox{-}&*(red!20)\mbox{-}&*(red!20)\mbox{-}\\
  &           \mbox{S}&           \mbox{N}&        \mbox{D0}    &    \mbox{D2}       & \mbox{D4}  \\
\end{ytableau}
}
\center SDR

\scalebox{0.7}
{
\begin{ytableau}
\mbox{D4}&*(green!20)\mbox{+}&*(green!20)\mbox{+}&*(green!20)\mbox{+}&*(green!20)\mbox{+}& \\
\mbox{D2}&*(green!20)\mbox{+}&*(green!20)\mbox{+}&*(green!20)\mbox{+}& &*(red!20)\mbox{-}\\
\mbox{D0}&*(green!20)\mbox{+}&*(green!20)\mbox{+}& &*(red!20)\mbox{-}&*(red!20)\mbox{-}\\
\mbox{N}&*(green!20)\mbox{+}& &*(red!20)\mbox{-}&*(red!20)\mbox{-}&*(red!20)\mbox{-}\\
\mbox{S}& &*(red!20)\mbox{-}&*(red!20)\mbox{-}&*(red!20)\mbox{-}&*(red!20)\mbox{-}\\
  &           \mbox{S}&           \mbox{N}&             \mbox{D0}      &    \mbox{D2}     & \mbox{D4}  \\ 
\end{ytableau}
}
\center SIR

\scalebox{0.7}
{
\begin{ytableau}
\mbox{D4}&*(red!20)\mbox{-}&*(red!20)\mbox{-}&*(red!20)\mbox{-}&*(red!20)\mbox{-}& \\
\mbox{D2}&*(yellow!20)\mbox{0}&*(yellow!20)\mbox{0}&*(red!20)\mbox{-}&  &*(green!20)\mbox{+}\\                                                 
\mbox{D0}&*(green!20)\mbox{+}&*(green!20)\mbox{+}&  &*(green!20)\mbox{+}&*(green!20)\mbox{+}\\
\mbox{N}&*(yellow!20)\mbox{0}&  &*(red!20)\mbox{-}&*(yellow!20)\mbox{0}&*(green!20)\mbox{+}\\
\mbox{S}&  &*(yellow!20)\mbox{0}&*(red!20)\mbox{-}&*(yellow!20)\mbox{0}&*(green!20)\mbox{+}\\
  &           \mbox{S}&           \mbox{N}&       \mbox{D0}       &    \mbox{D2}      &  \mbox{D4}  \\
\end{ytableau}
}
\center SAR
\end{multicols}
\label{table:dnn_sf} 
\end{table}

The average SDR, SIR, and SAR values of the separated vocal and accompaniment signals for the 30 test songs are reported in Fig. \ref{fig:sdrsir}. To plot this figure, the average of the vocal and accompaniment for each song was calculated as $(\mbox{SDR}_{\mbox{voc}}+\mbox{SDR}_{\mbox{acc}})/2$ for each model. The definitions of the models in Fig. \ref{fig:sdrsir} are as follows: model ``S'' is for using DNN-A for source separation without enhancement; model ``N'' is for using DNN-A for separation followed by NMF for enhancement as proposed in \cite{Williamson:14:tsaipqss}; models D0, D2, and D4 are for using DNN-A for separation followed by using DNN-B for enhancement with regularization parameter $\lambda=0,0.2$, and $0.4$ respectively. 
%
%As can be seen from Fig. \ref{fig:sdrsir}, the differences in SDR and SIR between separation with enhancement using DNN-A followed by DNN-B (D0, D2, D4) and separation using DNN-A only is higher than zero for most of the results. This means, using DNN-B for enhancement decreases the interference between the separated sources and distortions. For the case of using NMF for enhancement (model N), the difference is mostly above the zero for both SDR and SIR values but it is lower than using DNN-B in most cases. Model D0 gives the best SAR values.
%

The data shown in Fig. \ref{fig:sdrsir} were analysed using non-parametric statistical methods \cite{Simpson:16:eassmhdnpsm} to determine the significance of the effects of enhancing the separated sources. A pair of models are significantly different statistically if $P < 0.05$, Wilcoxon signed-rank test \cite{Wilcoxon:45:icrm} and Bonferroni corrected \cite{hochberg:87:mcp}. Table \ref{table:dnn_sf}, shows the significant differences between each pair of models in Fig. \ref{fig:sdrsir}. In this table, we denote the models in the rows as significantly better than the models in the columns using the sign ``+'', the cases with significantly worse as ``-'' and the cases without significant differences as ``0''. For example, Model D4 is significantly better than all other models in SIR and model D0 is significantly better than all other models in SAR. As can be seen from this table and Fig. \ref{fig:sdrsir}, model S is significantly worse than all other models for SDR and SIR, which means there is significant improvements due to using the second stage of enhancement compared to using DNN-A only for separation without enhancement (model S). Also, we can see significant improvements in SDR, SIR and some SAR values between the proposed enhancement methods using DNNs (models D0 to D4) compared to the enhancement method in \cite{Williamson:14:tsaipqss} using NMF (model N). This means that the proposed enhancement methods using DNN-B is significantly better than using NMF for enhancement. Model D0 achieves the highest SAR values and it is also significantly better in SDR and SIR than models S and N, which means that using DNN-B for enhancement even without discriminative learning ($\lambda=0$) still achieves good results compared with no enhancement (S) or using NMF for enhancement (N). The regularization parameter $\lambda$ in models D0 to D4 has significant impact on the results, and can be used as a trade-off parameter between achieving high SIR values verses SAR and vice versa. 

From the above analysis we can conclude that using DNN-B for enhancement improves the quality of the separated sources by decreasing the distortion (high SDR values) and interference (high SIR values) between the separated sources. Using discriminative learning for DNN-B improves the SDR and SIR results. Using DNN-B for enhancement gives better results than using NMF for most SDR, SIR, and SAR values.

The implementation of the separation and enhancement approaches in this paper is available at:
\url{http://cvssp.org/projects/maruss/discriminative/}
\section{Conclusion}
\label{sec:conc}
In this work, we proposed a new discriminative enhancement approach to enhance the separated sources after applying source separation. Discriminative enhancement was done using a deep neural network (DNN) to decrease the distortion and interference between the separated sources. To consider the interaction between the sources in the mixed signal, we proposed to enhance all the separated sources together using a single DNN. We enhanced the separated sources discriminatively by introducing a new cost function that decreases the interference between the separated sources. Our experimental results show that the proposed discriminative enhancement approach using DNN decreases the distortion and interference of the separated sources. In our future work, we will investigate the possibilities of using many stages of enhancement (multi-stages of enhancement).
\section*{ACKNOWLEDGMENT}
This work is supported by grants EP/L027119/1 and EP/L027119/2 from the UK Engineering and Physical Sciences Research Council (EPSRC).

\bibliographystyle{splncs03.bst}
%\footnotesize{
\bibliography{refs}
%\end{thebibliography}
%}
\end{document}